\documentclass[]{beilstein}

\usepackage{bm}
\usepackage{graphicx}
\usepackage{amsmath}
\usepackage{amsfonts}
\usepackage{bbold}
\usepackage{xspace}

\usepackage{color}
\definecolor{Blue}{rgb}{0.3,0.3,0.9}
\definecolor{Red}{rgb}{0.9,0.3,0.3}
\definecolor{Green}{rgb}{0.3,0.6,0.3}
\definecolor{Black}{rgb}{0.0,0.0,0.0}

\newcommand{\sgn}{\mathrm{sgn\,}}

\newcommand{\revision}[1]{\textcolor{Black}{{#1}}}

\begin{document}

\title{Robust midgap states in band-inverted junctions under electric and magnetic fields}

\author*{\'{A}lvaro\ D\'{\i}az-Fern\'{a}ndez}{alvaro.diaz@ucm.es}

\author{Natalia del Valle}

\author{Francisco Dom\'{\i}nguez-Adame}

\affiliation{GISC, Departamento de F\'{\i}sica de Materiales, Universidad Complutense, E-28040 Madrid, Spain}

\maketitle

\begin{abstract}
Several IV-VI semiconductor compounds made of heavy atoms, such as Pb$_{1-x}$Sn$_{x}$Te, may undergo band-inversion at the $L$ point of the Brillouin zone upon variation of their chemical composition. This inversion gives rise to topologically distinct phases, characterized by a change in a topological invariant. In the framework of the $\mathbf{k}\cdot\mathbf{p}$ theory, band-inversion can be viewed as a change of sign of the fundamental gap. A two-band model within the envelope-function approximation predicts the appearance of midgap interface states with Dirac cone dispersions in band-inverted junctions, namely, when the gap changes sign along the growth direction. We present a thorough study of these interface electron states in the presence of crossed electric and magnetic fields, the electric field being applied along the growth direction of a band-inverted junction. We show that the Dirac cone is robust and persists even if the fields are strong. In addition, we point out that Landau levels of electron states lying in the semiconductor bands can be tailored by the electric field. Tunable devices are thus likely to be realizable exploiting the properties studied herein.
\end{abstract}

\keywords{Crystalline topological insulators; electric and magnetic fields; Landau levels; midgap states}

\section{Introduction} \label{sec:intro}

In 1982 Thouless \emph{et al.}~\cite{Thouless82} made a connection between the quantum Hall effect and a topological invariant, the so-called first Chern number \cite{Kane05}. The fact that a quantum Hall system was insulating in the bulk but had a quantized conductivity on the surface could be related to the non-trivial topology of the band structure. In 2006, topology came up to stage once again with the theoretical prediction by Bernevig \emph{et al.}~\cite{Bernevig06} of a topological insulating behaviour in a HgTe/CdTe quantum well. The difference between the latter and the quantum Hall system lies in the fact that the quantum well required no magnetic field at all, but just relativistic corrections (Darwin and mass-velocity interactions) large enough so as to invert the $\Gamma_6$ and $\Gamma_8$ bands~\cite{Chu08}. The HgTe/CdTe quantum well possesses non-trivial edge states when a certain width is exceeded. In 2007, experiments verified this remarkable result and established the existence of the quantum spin Hall effect~\cite{Konig07}. However, no clear signatures of conductance quantization have been observed yet~\cite{Gusev11,Grabecki13}.

Besides II-VI compound semiconductors, such as HgTe, IV-VI semiconductors support non-trivial edges states as well~\cite{Hsieh12}. In this regard, Dziawa \emph{et al.} reported evidence of topological crystalline insulator states in Pb$_{1-x}$Sn$_{x}$Se~\cite{Dziawa12}. High resolution scanning tunneling microscopy studies of these topological crystalline insulators provided strong evidence of the coexistence of massless Dirac fermions, protected by crystal symmetry, with massive Dirac fermions consistent with crystal-symmetry breaking~\cite{Okada13}. Recently, this results have received further support with the aid of Dirac Landau level spectroscopy~\cite{Serbyn14,Phuphachong17}.

Band-inverted structures were already studied back in the $80$'s and $90$'s under the name of band-inverted junctions, in which the fundamental gap has opposite sign on each semiconductor. A salient feature is the existence of interface states lying within the gap, provided that the two gaps overlap (see Refs.~\cite{Volkov85,Korenman87,Agassi88,Pankratov90,Kolesnikov97} and references therein). These states are protected by symmetry, and are responsible for the conducting properties of the surface. In IV-VI heterojunctions, such as PbTe/SnTe, interface states are accurately described by means of a two-band model using the effective $\mathbf{k}\cdot\mathbf{p}$ approximation~\cite{Kriechbaum86,Ando15}. The equation governing the conduction- and valence-band envelope functions reduces to a Dirac-like equation after neglecting far-band corrections. In view of this analogy, exact solutions can be then straightforwardly found by means of supersymmetric~\cite{Pankratov90} or Green's function approaches~\cite{Adame94}. 
\revision{In the context of symmetry-protected topological phases, our model can be applied not only to topological crystalline insulators, like the ones mentioned above~\cite{Hsieh12}, but also to more general three-dimensional topological insulators in contact with a trivial insulator, such as Bi$_2$Se$_3$~\cite{Zhang12,Goerbig17}. In the former case, mirror symmetry makes it possible to define mirror Chern numbers, which determine the topological crystalline phase~\cite{Hsieh12}. In the latter, time-reversal symmetry, parity and particle-hole symmetry allow us to define a topological index given by the sign of the Dirac mass~\cite{Zhang12}.
}

\revision{In 1994, Agassi studied the case of a band-inverted junction with a magnetic field applied parallel to the junction~\cite{Agassi94}. This author showed that the Dirac point remains robust upon the application of a magnetic field of arbitrary strengths and that the Landau levels in the continuum split for non-zero values of the in-plane momentum in the direction perpendicular to the magnetic field. By means of the modern theory of symmetry-protected topological phases, the protection of the Dirac point can be understood in the case of topological crystalline insulators from the fact that a magnetic field perpendicular to a mirror plane renders a system that is still symmetric about that plane~\cite{Hsieh12}. This is not the case in a magnetic field parallel to the mirror plane, where the Dirac cone turns into the usual relativistic Landau levels~\cite{Volkov85,Volkov87,Agassi88}. Going back to the parallel magnetic field, Agassi demonstrated that for large values of this in-plane momentum, the states evolve to the bulk Landau states and the midgap state becomes the zero Landau level, usual of these Dirac systems. The reason is that the in-plane momentum perpendicular to the magnetic field is proportional to the center of the Landau orbits. If it becomes very large and the magnetic length is at the same time small (which happens for large magnetic fields), then the orbits do not intersect the junction and they might not notice that boundary. Hence, the case of most interest is in the vicinity of low in-plane momentum perpendicular to the field, where the states differ the most from the Landau levels of the bulk and we can see the effects of the interface.}

In this same topic of external fields applied to band-inverted junctions, we have recently studied band-inverted junctions based on IV-VI compounds using a two-band model when an electric field is applied along the growth direction~\cite{Diaz-Fernandez17a}. We have demonstrated that the Dirac cone of midgap states is robust against moderate values of the electric field but Fermi's velocity decreases quadratically with the applied field. The aim of this paper is to characterize electron states in band-inverted junctions using a two-band model in the presence of crossed magnetic and electric fields, the former parallel to the junction, the latter perpendicular to it. We show that the Dirac cone of midgap states arising in the single junction configuration is robust against crossed electric and magnetic fields. In addition, Landau levels of electron states lying in the semiconductor bands can be tailored by the electric field. Finally, the electronic structure of band-inverted junctions when the magnetic field is applied along the growth direction, parallel to the electric field, will also be briefly discussed for comparison.

\section{Theoretical model}   \label{sec:model}

\begin{figure}
\centerline{\includegraphics[width=0.5\columnwidth]{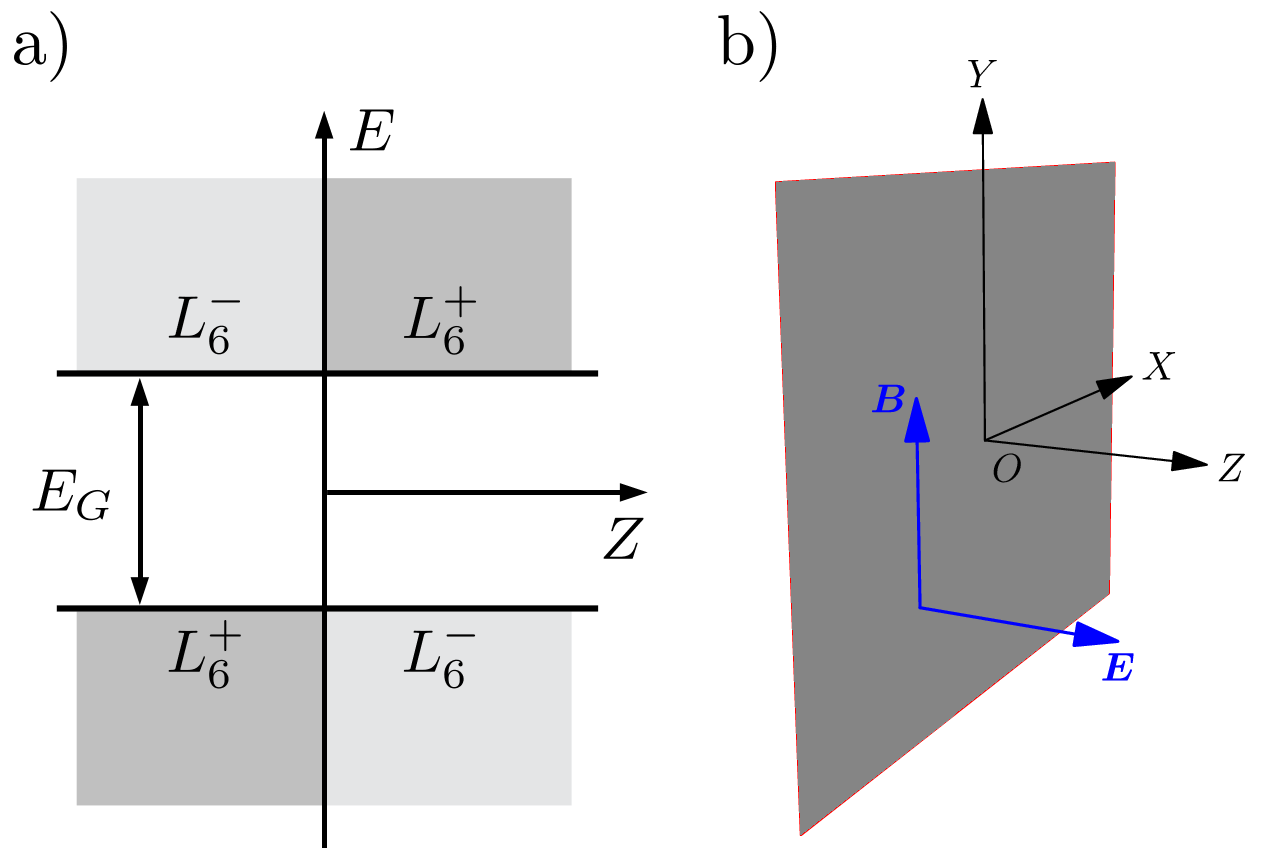}}
\caption{(a)~$L_6^{+}$ and $L_6^{-}$ band-edge profile of an abrupt band-inverted junction with aligned and same-sized gaps, located at the $XY$ plane, and b)~schematic view of the applied electric and magnetic fields.}
\label{fig1}
\end{figure}

We consider heterojunctions of IV-VI compound semiconductors, such as Pb$_{1-x}$Sn$_{x}$Te and Pb$_{1-x}$Sn$_{x}$Se. The latter are known to shift from being semiconductors to topological crystalline insulators due to the band inversion at the $L$ points of the Brillouin zone as the Sn fraction increases~\cite{Hsieh12,Assaf16, Hasan12}. In order to keep the algebra as simple as possible, we restrict ourselves to the symmetric heterojunction with same-sized and aligned gaps, as depicted in Figure~\ref{fig1}(a). This assumption simplifies the calculations while keeping the underlying physics~\cite{Diaz-Fernandez17b}. Thus, a single and abrupt interface presents the following profile for the magnitude of the gap 
\begin{equation}
E_{G}(z)=2\Delta\,\sgn(z)\ ,
\label{eq01}
\end{equation}
where $\sgn(z)=|z|/z$ is the sign function. Here $Z$ axis is parallel to the growth direction $[111]$. 

The envelope functions of the electron states near the band extrema $L_{6}^{+}$ and $L_{6}^{-}$ in IV-VI compounds are determined from the following Dirac-like Hamiltonian~\cite{Agassi88,Pankratov90,Ando15} 
\begin{equation}
\mathcal{H}_{0}=v\,{\bm\alpha}\cdot\mathbf{p}+\frac{1}{2}\,E_{G}(z)\,\beta\ .
\label{eq02}
\end{equation}
Here ${\bm\alpha}=(\alpha_x,\alpha_y,\alpha_z)$ and $\beta$ denote the usual $4\times 4$ Dirac matrices, $\alpha_{i}=\sigma_x\otimes \sigma_{i}$ and $\beta=\sigma_z\otimes\mathbb{1}_2$, $\sigma_i$ and $\mathbb{1}_n$ being the Pauli matrices and $n\times n$ identity matrix, respectively. Moreover, $v$ is an interband matrix element having dimensions of velocity and it is assumed scalar, corresponding to isotropic bands around the $L$ point. 
\revision{It is worth mentioning that the bands of IV-VI compounds around the $L$ points are actually anisotropic. Nevertheless, this anisotropy can be absorbed in the definition of the dimensionless parameters defined below. That is, it is possible to consider a direction-dependent velocity, but it will not change the results shown below, except for a proportionality constant in the definition of the dimensionless in-plane momenta (see Refs.~\cite{Ando15,Diaz-Fernandez17b} for further details).}
In addition, we focus on states close to one of the $L$ points of the Brillouin zone~\cite{Hsieh12} and neglect other valleys in what follows since midgap states are stable against gap opening by valley mixing. Also notice that only linear momentum terms are taken into account in equation~(\ref{eq02}) but quadratic momentum terms could have an impact of the electronic levels~\cite{Kriechbaum84,Buczko12}. However, the two-band model Hamiltonian~(\ref{eq02}) successfully describes the hybridization of interface states in band-inverted quantum wells~\cite{Diaz-Fernandez17c}, in perfect agreement with more elaborated models including quadratic momentum terms~\cite{Buczko12}.

The Hamiltonian~(\ref{eq02}) acts upon the envelope function ${\bm\chi}(\mathbf{r})$, which is a four-component vector composed of the two-component spinors ${\bm\chi}_{+}(\mathbf{r})$ and ${\bm\chi}_{-}(\mathbf{r})$ belonging to the $L_{6}^{+}$ and $L_{6}^{-}$ bands. The interface momentum is conserved and the envelope function can be expressed as ${\bm\chi}(\mathbf{r})=\widetilde{\bm\chi}(z)\exp(i\,\mathbf{r}_{\bot}\cdot\mathbf{k}_{\bot})$, where it is understood that the subscript $\bot$ in a vector indicates the nullification of its $z$--component. In the case of aligned and same-sized gaps, it is found that $\widetilde{\bm\chi}(z)\sim \exp(-|z|/d)$, with $d=\hbar v/\Delta$ and the interface dispersion relation is a single Dirac cone $E(\mathbf{k}_{\bot})=\pm \hbar v|\mathbf{k}_{\bot}|$, where the origin of energy is taken at the center of the gaps~\cite{Adame94}. $v$ is the group velocity at the Fermi level in undoped samples and it will be referred to as Fermi velocity hereafter.

\section{Electron states under crossed electric and magnetic fields}   \label{sec:single}

We now turn to the electronic states of a single band-inverted junction subjected to a perpendicular electric field $\mathbf{F}=F\,\widehat{\bm z}$ and a parallel magnetic field $\mathbf{B}=B\,\widehat{\bm y}$, as shown schematically in Figure~\ref{fig1}(b). By choosing the Landau gauge, the vector potential is given as $\mathbf{A}(z)=Bz\,\widehat{\bm x}$.

The electrostatic potential $eFz$ and the vector potential $\mathbf{A}(z)$ only depend on the $z$--coordinate. Therefore, $\mathbf{p}_\bot=\hbar \mathbf{k}_\bot$ is a constant of motion and the envelope function can still be factorized to the form ${\bm\chi}(\mathbf{r})=\widetilde{\bm\chi}(z)\exp(i\,\mathbf{r}_{\bot}\cdot\mathbf{k}_{\bot})$. Now the longitudinal envelope function $\widetilde{\bm\chi}(z)$ satisfies the following Dirac equation
\begin{equation}
\big[\mathcal{H}_{0}+ev\,{\bm\alpha}\cdot\mathbf{A}(z)+eFz-E\big]
\widetilde{\bm\chi}(z)=0\ ,
\label{eq03}
\end{equation}
where $\mathcal{H}_{0}$ is given by~(\ref{eq02}). To address this problem we shall follow the Feynman-Gell-Mann \emph{ansatz}~\cite{Feynman58} and define a new four-component vector ${\bm \psi}(z)$ as
\begin{equation}
\widetilde{\bm\chi}(z)=\big[\mathcal{H}_{0}+ev\,{\bm\alpha}\cdot\mathbf{A}(z)-eFz+E\big]{\bm\psi}(z)\ .
\label{eq04}
\end{equation}
It is convenient to introduce the following dimensionless quantities ${\bm\kappa}_{\bot}=\mathbf{k}_{\bot}d$, $\xi=z/d$, $\varepsilon=E/\Delta$, $f=eFd/\Delta$, and $b=eBd^2/\hbar$. Notice that $f/2$ is the ratio between the electric potential drop across the spatial extent of the midgap states $d=\hbar v/\Delta$ in the absence of fields and the magnitude of the fundamental gap $2\Delta$. Similarly, $b$ is the square of the ratio between $d$ and the magnetic length $\ell=\sqrt{\hbar/eB}$. Hereafter we shall consider $b>f\geq 0$ for reasons that will become clear shortly. Let us define
\begin{align}
p&=\frac{1}{2\mu^{2}}\,
\left(\epsilon^2-{\bm\kappa}_{\bot}^2-1+\frac{(\kappa_xb+\epsilon f)^2}{\mu^4}\right) \ ,
\nonumber \\
s&=-\sqrt{2}\mu\left(\xi+\frac{\kappa_xb+\epsilon f}{\mu^4}\right) \ ,
\label{eq05}
\end{align}
where $\mu=\left(b^2-f^2\right)^{1/4}$ is real. Then, inserting the \emph{ansatz}~\ref{eq04} in Equation~\ref{eq03} and taking into account Equation~\ref{eq05}, we get
\begin{equation}
\left[-\frac{d^2}{ds^2}+\frac{s^2}{4}-p+\mathcal{M}\right]{\bm\psi}(s)=
\delta\left(s-s_0\right)\mathcal{N}{\bm\psi}(s) \ ,
\label{eq06}
\end{equation}
where $s_0\equiv s(\xi=0)$. Here $\mathcal{M}$ and $\mathcal{N}$ are $4\times 4$ matrices given by
\begin{equation}
\mathcal{M}=\frac{i}{2\mu^2}\,(b\alpha_x+f)\,\alpha_z \ ,
\qquad
\mathcal{N}=i\,\frac{\sqrt{2}}{\mu}\,\alpha_z\,\beta \ .
\label{eq07}
\end{equation}
Let us diagonalize the left-hand side of the equation by introducing a unitary matrix $U$ such that $\mathcal{M}=U(\beta/2) U^{-1}$. Doing so and defining $\mathcal{W}=U^{-1}\mathcal{N}U$ and ${\bm\phi}(s)=U^{-1}{\bm\psi}(s)$ we obtain
\begin{equation}
\left[-\frac{d^2}{ds^2}+\frac{s^2}{4}-p+\frac{1}{2}\,\beta\right]{\bm\phi}(s)=
\delta\left(s-s_0\right)\mathcal{W}{\bm\phi}(s) \ .
\label{eq08}
\end{equation}
In order to solve Equation~\ref{eq08} we shall use the Green's function method. The solution to Equation~\ref{eq08} will be given by
\begin{align}
{\bm\phi}(s)&=\int_{-\infty}^{\infty}ds^{\prime}~G(s,s^{\prime})\delta(s^{\prime}-s_0)\mathcal{W}{\bm\phi}(s^{\prime})\nonumber\\
&=G(s,s_0)\mathcal{W}{\bm\phi}(s_0) \ ,
\label{eq09}
\end{align}
where the retarded Green's function $G(s,s^{\prime})$ satisfies
\begin{equation}
\left[-\frac{\partial^2}{\partial s^2}+\frac{s^2}{4}-p+\frac{1}{2}\beta\right]G(s,s^{\prime})=
\delta\left(s-s^{\prime}\right)\mathbb{1}_4 \ ,
\label{eq10}
\end{equation}
and $G(s,s^{\prime})\to 0$ as $|s|,|s^{\prime}|\to\infty$. Notice that $G(s,s^{\prime})$ is continuous on the line $s=s^{\prime}$. Equation~~(\ref{eq09}) can be particularized for $s=s_0$, leading to a homogeneous system of equations with non-trivial solutions existing for energies satisfying the vanishing of the determinant
\begin{equation}
 \det\left[\mathbb{1}_4-G(s_0,s_0)\mathcal{W}\right]=0 \ .
 \label{eq11}
\end{equation}
Since $G(s,s^{\prime})$ can be considered as the inverse of the operator that acts upon it and the latter is diagonal, we may consider $G(s,s^{\prime})$ to be block diagonal. Hence,
\begin{equation}
G(s,s^{\prime})=\begin{pmatrix}
g_{+}(s,s^{\prime})\mathbb{1}_2 & \mathbb{0}_2 \\ 
\mathbb{0}_2 & g_{-}(s,s^{\prime})\mathbb{1}_2 
\end{pmatrix} \ ,
 \label{eq12}
\end{equation}
where $\mathbb{0}_2$ is the $2\times 2$ null matrix and the scalar functions $g_{\pm}(s,s^{\prime})$ satisfy
\begin{equation}
\left[-\frac{\partial^2}{\partial s^2}+\frac{s^2}{4}-p_{\pm}\right]g_{\pm}(s,s^{\prime})=
\delta\left(s-s^{\prime}\right) \ ,
\label{eq13}
\end{equation}
with $p_{\pm}=p\mp 1/2$. Since $s$ is real because we have chosen $\mu$ to be so, then $s^2>0$ and this equation corresponds to a harmonic oscillator. Notice that this would not be the case if $\mu$ were imaginary as in that case $s^2<0$ and we would not have the positive parabola required for a harmonic oscillator. The solution to this problem is known to be~\cite{Adame91,Glasser15}
\begin{equation}
 g_{\pm}(s,s^{\prime})=\frac{1}{\sqrt{2\pi}}\,\Gamma\left(\frac{1}{2}-p_{\pm}\right)D_{p_{\pm}-1/2}(s_{>})D_{p_{\pm}-1/2}(-s_{<}) \ ,
 \label{eq14}
\end{equation}
where $\Gamma(z)$ is the Gamma function, $D_{\gamma}(z)$ is the parabolic-cylinder function, $s_{>}=\max(s,s')$ and $s_{<}=\min(s,s')$. Now that we have $G(s,s')$, it is straightforward to obtain from~(\ref{eq11}) that $g_{+}(s_0,s_0)g_{-}(s_0,s_0)=\mu^2/2$. Equivalently
\begin{equation}
  D_p(s_0)D_p(-s_0)D_{p-1}(s_0)D_{p-1}(-s_0)+\frac{\pi\mu^2}{p\Gamma^2(-p)}=0 \ .
  \label{eq15}
\end{equation}
Equation~(\ref{eq15}) determines the dispersion relation $\epsilon({\bm \kappa})$ of midgap interface states as well as Landau levels lying in the semiconductor bands. \revision{It reduces to the result found by Agassi when the electric field vanishes~\cite{Agassi94}.}

\section{Results and Discussions}

\subsection{Energy levels in the absence of electric field}

\revision{This section is added for completeness and essentially reproduces the results found by Agassi~\cite{Agassi94} for small values of $\kappa_x$. However, we will be able to give approximate dispersion relations for the midgap state and the Landau levels which will provide us with a clearer view of the effect of the magnetic field in our case of interest. This section then corresponds to the $f=0$ case, where approximate results can be obtained.} In fact, these results are exact when $\kappa_x=0$, where $s_0=0$. Let us explore this last case. Using $\Gamma(1+z)=z\Gamma(z)$ and the Legendre duplication formula $\Gamma(2z)=2^{2z-1}\Gamma(z)\Gamma(z+1/2)/\sqrt{\pi}$, it is straightforward to obtain from Equation~\ref{eq15}
\begin{equation}
 \frac{1+2p\mu^2}{p^2\Gamma^2(-p)}=0 \ .
 \label{eq16}
\end{equation}
There are now two possibilities, either the numerator goes to zero or the denominator goes to infinity. If $p<0$, it is necessary to have numerator equal to zero, which amounts to having,
\begin{equation}
 \varepsilon=\pm \kappa_y \ .
 \label{eq17}
\end{equation}
This is nothing but a Dirac linear dispersion in the $y$-direction. It is remarkable that the Dirac point remains robust for any strength of $b$. Taking into account the definition of $p$, the case where $p<0$ corresponds to $|\varepsilon|<1$ at $\kappa_x=\kappa_y=0$, meaning that these states lie within the gap.

Let us explore other possibilities. If $p=0$, then both the numerator and the denominator are finite, which implies that $p=0$ is not a solution. The other option where $p>0$ is only achieved if the denominator goes to infinity because the numerator is always positive in this case. For this to happen, $p$ must be a positive integer. The corresponding energies are the usual Landau levels of a relativistic particle
\begin{equation}
 \varepsilon=\pm \sqrt{1+2nb+\kappa_y^2} \ ,
 \qquad
 n=1,2,\cdots
 \label{eq18}
\end{equation}
There is no zero Landau level because the requirement of $p>0$ implies $|\varepsilon|>1$ at $\kappa_x=\kappa_y=0$, that is, Landau levels live outside the gap.

With this results in mind, we can now turn to the case where $\kappa_x\neq 0$, but $s_0\to 0$. After some tedious algebra we arrive to the following expression
\begin{equation}
 \frac{1}{p^2\Gamma^2(-p)}\left\{1+2p(\mu^2-s_{0}^{2})+\left[\eta(p)+\frac{1}{\eta(p)}\right]\right\} = 0 \ ,
 \label{eq19}
\end{equation}
where
\begin{equation}
 \eta(p)=\frac{p}{2}\left[\frac{\Gamma(-p/2)}{\Gamma(1/2-p/2)}\right]^2 \ .
 \label{eq20}
\end{equation}

Notice that if $s_0=0$ we obtain back Equation~\ref{eq16}, corresponding to $\kappa_x=0$. Now if $\kappa_x\neq 0$, then either the term in curly brackets is zero or the prefactor multiplying this term is zero. As before, if the prefactor is zero then $p$ is a positive integer. However, that would imply two possible energies for each integer, but numerically we will show briefly that this is not the case. Thus, we must consider the term in curly brackets to be equal to zero. If we consider $b\to 0$, but at the same time $\kappa_x\to 0$ sufficiently fast so that $s_0\to 0$, then it is not difficult to obtain for the states inside the gap
\begin{equation}
 \varepsilon=\pm\sqrt{\left(1-\frac{5b^2}{4}\right)\kappa_x^2+\kappa_y^2} \ ,
 \label{eq21}
\end{equation}
whereas for the Landau levels we obtain to lowest order in $\kappa_x$
\begin{equation}
 \varepsilon = \pm \sqrt{1+2nb+\kappa_y^2\pm\sqrt{\frac{8nbc(n)}{1+2nb}}\,\kappa_x} \ ,
 \label{eq22}
\end{equation}
where $c(n)$ results from the expansion around integer values of $p$ of $\eta(p)+\eta^{-1}(p)+2\approx c(n)(p-n)^{-2}$. For instance, $c(1)=2/\pi$, $c(2)=1/\pi$, $c(3)=3/2\pi,\dots$ Before we consider each case, it is important to mention that the approximation of low $b$ corresponds to the range of interest in experiments since typically $d\approx 4.5$~nm and as a result $b = 0.5$ corresponds to a very large magnetic field of about $16\,$T. 

\begin{figure}
\centerline{\includegraphics[width=0.30\columnwidth]{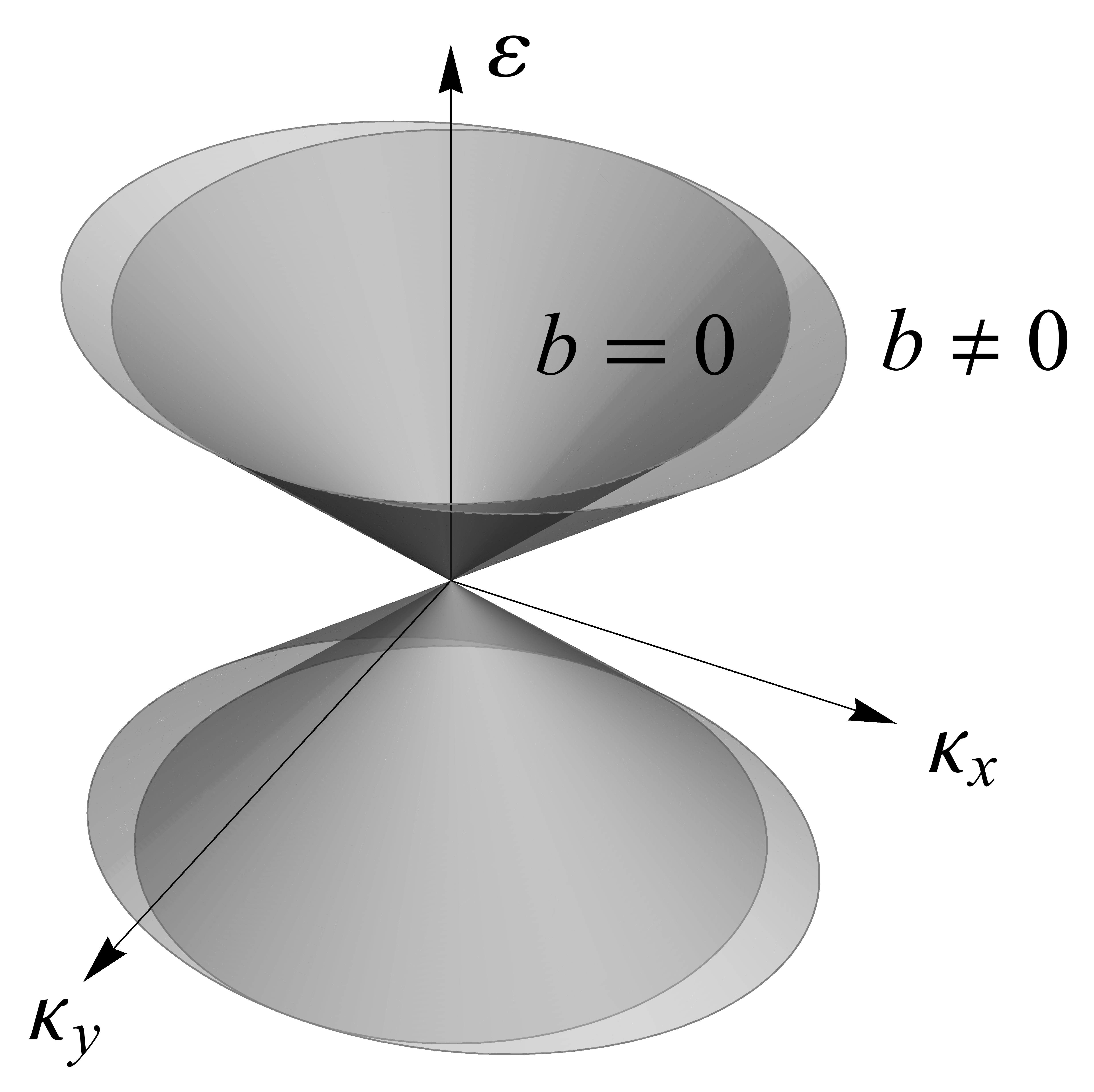}}
\caption{Dirac cones with, $b\neq 0$, and without, $b=0$, a magnetic field applied. The original cone is distorted along the $x$-direction and the Fermi velocity, i.e. the slope, becomes anisotropic.}
\label{fig2}
\end{figure}

\begin{figure}
\centerline{\includegraphics[width=0.5\columnwidth]{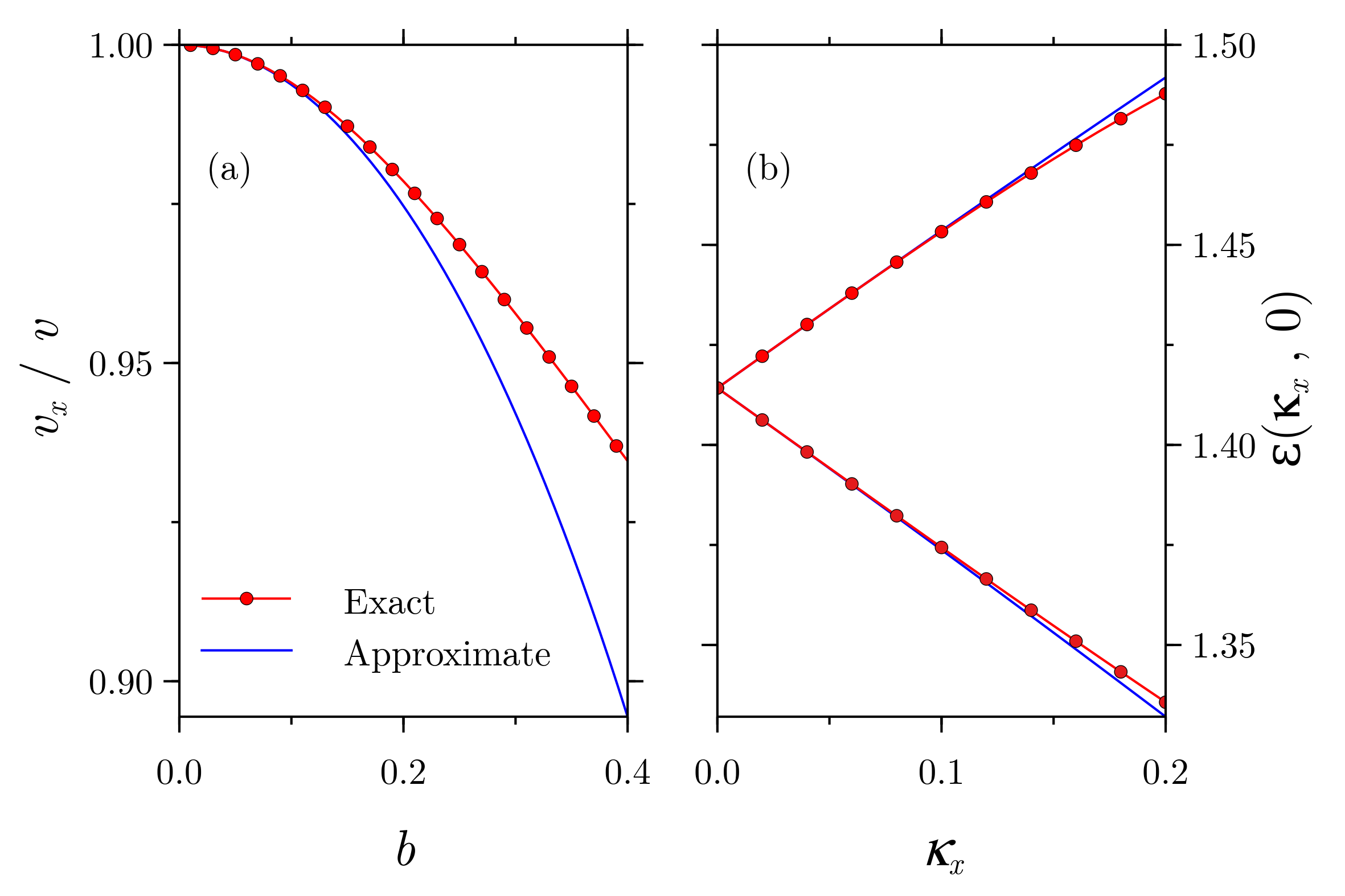}}
\caption{Comparison between exact and approximate results given by (a)~Equation~\ref{eq21} and (b)~Equation~\ref{eq22}. In panel~(a) the Fermi velocity along the $x$-direction, calculated as the slope of the dispersion relation, is substantially reduced and the agreement between the exact and approximate results is noteworthy up to $b\lesssim 0.2$. In panel~(b), the Landau level splitting in the $x$-direction is very well predicted even for $b=0.5$, as shown for the first level.}
\label{fig3}
\end{figure}

Let us now consider each case separately. On the one hand, Equation~\ref{eq21} corresponds to an elliptic cone and for $b=0$ we recover the original Dirac cone. It is not only remarkable, as we mentioned above, the fact that the Dirac point is robust, but also that the shape of the dispersion relation remains a cone but slightly widened in the $x$-direction, as shown in Figure~(\ref{fig2}). Hence, the Fermi velocity becomes anisotropic and can actually be modulated with the magnetic field. It is expected that the application of an electric field will lead to further reduction of the Fermi velocity~\cite{Diaz-Fernandez17a}. We will prove later that this is actually the case. 
In Figure~(\ref{fig3}a) we show a comparison between the Fermi velocity in the $x$-direction (recall that it does not change in the $y$-direction) given by the numerical evaluation of~(\ref{eq15}) and the approximation~(\ref{eq21}). The agreement is noteworthy for low values of $b$. 

We can now focus on the Landau levels given by Equation~\ref{eq22}. As it can be seen, for non zero values of $\kappa_x$, each Landau level at $\kappa_x=0$ splits into two Landau levels at $\kappa_x\neq 0$ due to the occurrence of a $\pm$ sign inside the square root. The comparison for the first Landau level, $n=1$, between the approximate result and the numerical calculations from Equation~\ref{eq15} are shown in Figure~\ref{fig3}(b). In contrast to Figure~\ref{fig3}(a), there is still agreement between both approaches for a large field of $b=0.5$.
 
\subsection{Energy levels at finite electric field}

\begin{figure}
\centerline{\includegraphics[width=0.5\columnwidth]{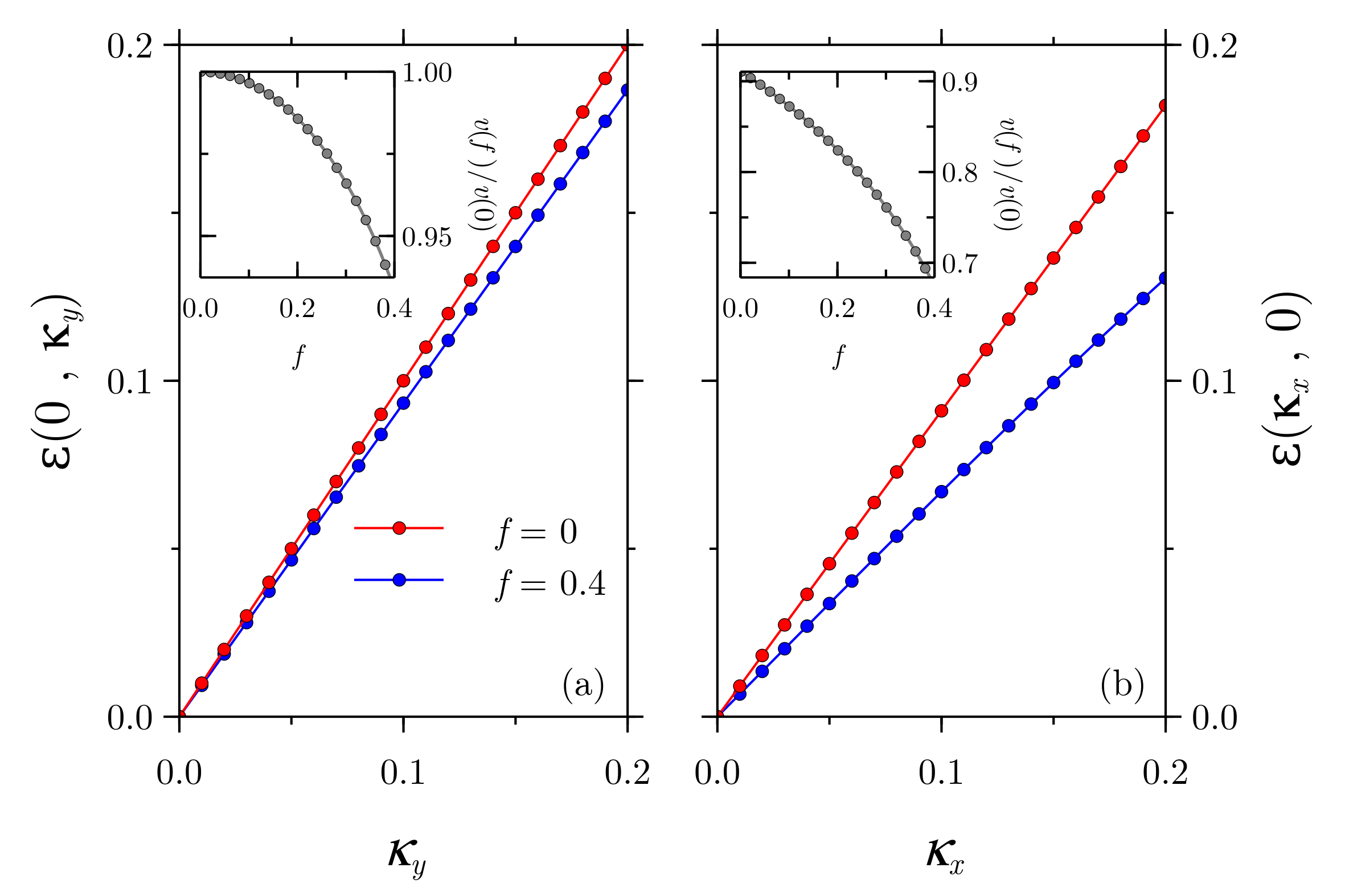}}
\caption{The additional effect of the electric field leads to a further reduction of the Fermi velocity in the $x$-direction and to a reduction along the $y$-direction as well. The Dirac point, however, remains robust. The insets show the Fermi velocity reduction as a function the electric field for a fixed magnetic field of $b=0.5$.}
\label{fig4}
\end{figure}

Let us now draw our attention to the $f\neq 0$ case. In contrast to the $f=0$ case, we have been unable to obtain explicit expressions of the dispersion relation, but the numerics shows remarkable results. Let us focus first on the midgap states. Since the magnetic field did not erase the Dirac point and based on known results of a band-inverted junction under an electric field~\cite{Diaz-Fernandez17a,Diaz-Fernandez17b}, it seems plausible to argue that the effect of the electric field will be to enhance the reduction of the Fermi velocity in the $x$-direction and introduce a reduction in the $y$-direction as well, leaving however the Dirac point untouched. That is indeed what we observe and we show our results in Figure~(\ref{fig4}). The insets show the Fermi velocity reduction as a function of the electric field for a fixed $b=0.5$. It is remarkable how the Fermi velocity along the $x$-direction is substantially decreased in band-inverted junctions subject to crossed magnetic and electric fields. 

\begin{figure}
\centerline{\includegraphics[width=0.45\columnwidth]{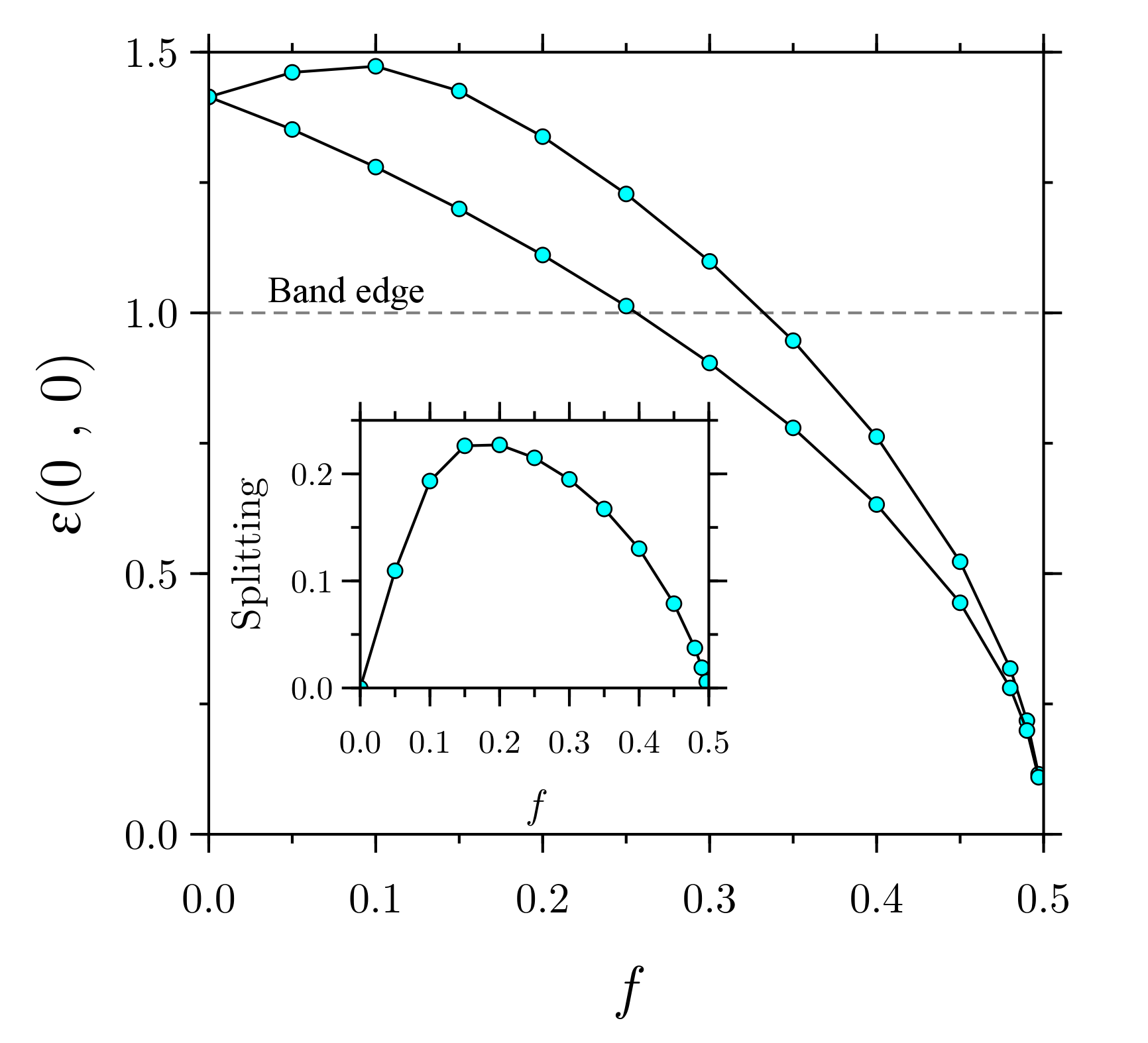}}
\caption{Splitting of the Landau levels at $\mathbf{k}_{\bot}=0$ and $b=0.5$ as a function of the electric field. It is important to notice that the Landau levels move below the band edge due to the bending by the electric field (see main text for details).}
\label{fig5}
\end{figure}

We may now turn to the evolution of the Landau levels as a function of the electric field. For simplicity, we shall consider only the first Landau level. It is illustrative to consider first the evolution of the lowest point of the Landau bands, that is, ${\bm\kappa}_{\bot}=0$. If the electric field is zero, we already know what the energy will be from the discussion above. However, as we turn on the electric field, a splitting similar to the one we had with $\kappa_x$ begins to develop. That splitting increases with electric field, up to a point where it starts decreasing again as $f$ approaches $b$. In the limiting case where $f\to b$, the splitting goes to zero, as we show in Figure~(\ref{fig5}) for $b=0.5$.

In Figure~(\ref{fig5}) it may be surprising to see that the Landau bands shift below the band edge, leading to the apparent and erroneous belief that the latter enter the band gap. The effect of the electric field is to bend the constant band edges shown in Figure~(\ref{fig1}a) upwards due to the presence of the electrostatic potential $eFz$, and so the Landau levels of the conduction band can move towards lower energies as long as the corresponding wave functions are not inside the band gap in position space.

\begin{figure}
\centerline{\includegraphics[width=0.5\columnwidth]{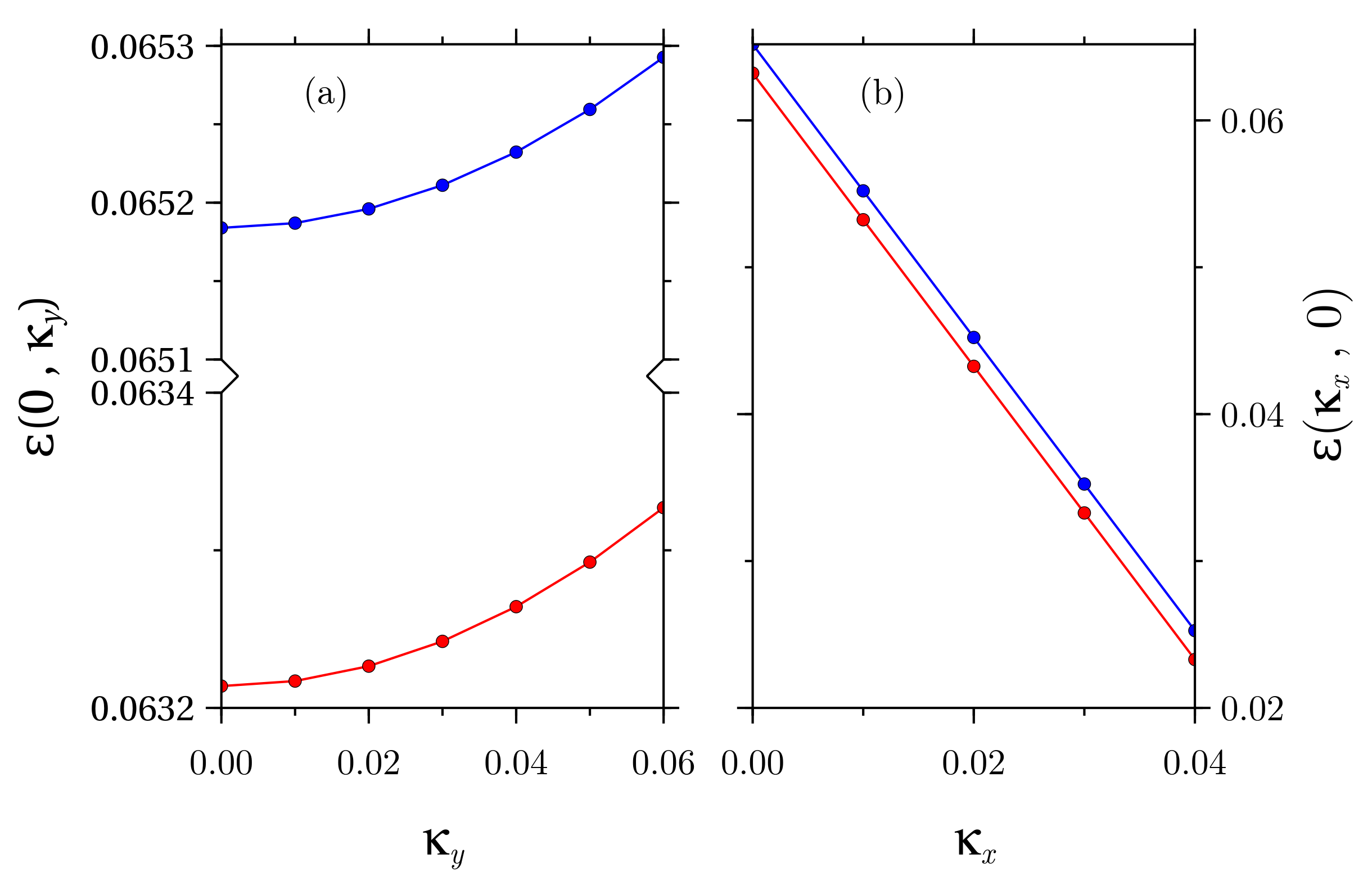}}
\caption{First Landau level dispersions for $b=0.5$ and $f=0.499$. In (a),~the original parabolic dispersion along the $y$-direction splits into two parabolic dispersions with energies below the band edge for the chosen fields, whereas in (b)~the previously obtained splitting in the $x$-direction is now exclusively downwards.}
\label{fig6}
\end{figure}

Finally, it deserves consideration the previous discussion for low values of ${\bm\kappa}_{\bot}$. As we can see in Figure~(\ref{fig6}), the parabolic dispersion that we obtained in the $y$-direction in the absence of an electric field splits into two parabolic bands. However, it is more remarkable to see that, instead of obtaining a splitting similar to that in Figure~(\ref{fig3}b), the dispersion goes downwards.

\section{Electron states under perpendicular  electric and magnetic fields}   \label{sec:parallel}

In previous sections we considered electron states when the magnetic field is parallel to the band-inverted junction, as depicted in Figure~\ref{fig1}. For completeness, we now briefly discuss the salient features of the energy spectrum when the electric and magnetic
fields are both perpendicular to the junction. The vector potential is now given as $\mathbf{A}(x)=Bx\,\widehat{\bm y}$ in the Landau gauge and thus $\mathbf{B}=B\,\widehat{\bm z}$. Starting from the Dirac equation~(\ref{eq03}) with this vector potential and using the Feynman-Gell-Mann \emph{ansatz}~(\ref{eq04}), one is led to a two-dimensional Schr\"{o}dinger equation in the $XZ$ plane. The resulting equation turns out to be separable in the $x$ and $z$ coordinates and can be straightforwardly solved by Green's function techniques. At low or moderate electric and magnetic fields ($f<b<1$), the energy levels within the gap are found to be
\begin{equation}
 \varepsilon = \pm \sqrt{2nb}\,\left(1-\frac{5}{8}\,f^2\right) \ ,
 \label{eq23}
\end{equation}
where $n=0,1,\ldots$ The above expression resembles the Landau levels of relativistic particles for an effective dimensionless magnetic field $b_\mathrm{eff}\equiv b(1-5f/8)^2\simeq b(1-5f/4)$. Therefore, the electric field decreases the Landau level spacing as in the previous field configuration.
There is yet another way of interpreting this result. If we undo the change of variables, we obtain for the energy the usual expression for the Landau levels that develop from a Dirac cone, the same as in graphene,
\begin{equation}
 E = \pm v_F(F)\sqrt{2eB\hbar n} \ ,
 \label{eq24}
\end{equation}
but with a renormalized Fermi velocity,
\begin{equation}
 v_F(F)\equiv v\left(1-\frac{5F^2}{8F_{\text{C}}^2}\right) \ ,
 \label{eq25}
\end{equation}
where $F_{\text{C}}=\Delta/ed$. In Ref.~\cite{Diaz-Fernandez17a}, it was anticipated that this renormalization of the Fermi velocity in a band-inverted junction with a perpendicular electric field could be measured by means of magnetotransport experiments, a prediction that is confirmed here.

\section{Conclusions}

In conclusion, we have studied band-inverted junctions under crossed electric and magnetic fields, the electric field being applied along the growth direction. Electron states are described by a spinful two-band model that is equivalent to the Dirac model for relativistic electrons. The mass term is half the bandgap and changes its sign across the junction. For the sake of algebraic simplicity we assumed same-sized and aligned gaps, although this is not a serious limitation to the validity of the results~\cite{Diaz-Fernandez17b}.

In the absence of external fields, it is well known that band-inverted junctions support topologically protected states located at the interface. Their energy lies within the common gap of the two semiconductors and the dispersion relation is a Dirac cone~\cite{Volkov85,Agassi88,Pankratov90,Adame94}. The Dirac cone remains even if an electric field perpendicular to the junction is applied, but it widens and the Fermi velocity is quadratically reduced with the electric field~\cite{Diaz-Fernandez17a,Diaz-Fernandez17b}. In this paper we have proved that electrons with energy within the gap still behave as massless fermions when an additional magnetic field parallel to the band-inverted junction is applied. The original Dirac cone widens only in the direction perpendicular to the magnetic field but remarkably the dispersion relation remains gapless. Hence the Fermi velocity becomes anisotropic and the combination of both electric and magnetic fields allows the Fermi velocity to be finely tuned. In addition, states lying within the semiconductor bands display relativistic-like Landau levels that split upon the application of the magnetic and electric fields. Interestingly, if both fields are parallel to the growth direction, the Landau level spacing can be further reduced by the electric field. We expect that the control of the Fermi velocity of topologically protected states will have applications for the design of novel electronic devices based on topological materials.

\begin{acknowledgements}
The authors thank L. Chico and J. W. Gonz\'{a}lez for helpful discussions. This work was supported by the Spanish MINECO under grant MAT2016-75955. 
\end{acknowledgements}

\bibliography{references}

\begin{thebibliography}{34}

\setboolean{nobreakdashused}{false}\bibitem[Thouless et~al.(1982)Thouless,
  Kohmoto, Nightingale, and den Nijs]{Thouless82}
Thouless,~D.~J.; Kohmoto,~M.; Nightingale,~M.~P.; den Nijs,~M. \emph{Phys. Rev.
  Lett.} \textbf{1982}, \emph{49}, 405.


\setboolean{nobreakdashused}{false}\bibitem[Kane and Mele(2005)]{Kane05}
Kane,~C.~L.; Mele,~E.~J. \emph{Phys. Rev. Lett.} \textbf{2005}, \emph{95},
  146802.


\setboolean{nobreakdashused}{false}\bibitem[Bernevig et~al.(2006)Bernevig,
  Hughes, and Zhang]{Bernevig06}
Bernevig,~B.~A.; Hughes,~T.~L.; Zhang,~S.-C. \emph{Science} \textbf{2006},
  \emph{314}, 1757.


\setboolean{nobreakdashused}{false}\bibitem[Chu and Sher(2008)]{Chu08}
Chu,~J.; Sher,~A. \emph{Physics and Properties of Narrow Gap Semiconductors};
\newblock Springer: New York, 2008;
\newblock p 157.


\setboolean{nobreakdashused}{false}\bibitem[K{\"o}nig et~al.(2007)K{\"o}nig,
  Wiedmann, Br{\"u}ne, Roth, Buhmann, Molenkamp, Qi, and Zhang]{Konig07}
K{\"o}nig,~M.; Wiedmann,~S.; Br{\"u}ne,~C.; Roth,~A.; Buhmann,~H.;
  Molenkamp,~L.~W.; Qi,~X.-L.; Zhang,~S.-C. \emph{Science} \textbf{2007},
  \emph{318}, 766.


\setboolean{nobreakdashused}{false}\bibitem[Gusev et~al.(2011)Gusev, Kvon,
  Shegai, Mikhailov, Dvoretsky, and Portal]{Gusev11}
Gusev,~G.~M.; Kvon,~Z.~D.; Shegai,~O.~A.; Mikhailov,~N.~N.; Dvoretsky,~S.~A.;
  Portal,~J.~C. \emph{Phys. Rev. B} \textbf{2011}, \emph{84}, 121302.


\setboolean{nobreakdashused}{false}\bibitem[Grabecki et~al.(2013)Grabecki,
  Wr\'obel, Czapkiewicz, Cywi\ifmmode~\acute{n}\else \'{n}\fi{}ski,
  Giera\l{}towska, Guziewicz, Zholudev, Gavrilenko, Mikhailov, Dvoretski,
  Teppe, Knap, and Dietl]{Grabecki13}
Grabecki,~G.; Wr\'obel,~J.; Czapkiewicz,~M.; Cywi\ifmmode~\acute{n}\else
  \'{n}\fi{}ski,~L.; Giera\l{}towska,~S.; Guziewicz,~E.; Zholudev,~M.;
  Gavrilenko,~V.; Mikhailov,~N.~N.; Dvoretski,~S.~A.; Teppe,~F.; Knap,~W.;
  Dietl,~T. \emph{Phys. Rev. B} \textbf{2013}, \emph{88}, 165309.


\setboolean{nobreakdashused}{false}\bibitem[Hsieh et~al.(2012)Hsieh, Lin, Li,
  Duan, Bansil, and Fu]{Hsieh12}
Hsieh,~T.~H.; Lin,~H.; Li,~J.; Duan,~W.; Bansil,~A.; Fu,~L. \emph{Nat. Commun.}
  \textbf{2012}, \emph{3}, 982.


\setboolean{nobreakdashused}{false}\bibitem[Dziawa et~al.(2012)Dziawa,
  Kowalski, Dybko, Buczko, Szczerbakow, Szot, \L{}usakowska, Balasubramanian,
  Wojek, Berntsen, Tjernberg, and Story]{Dziawa12}
Dziawa,~P.; Kowalski,~B.~J.; Dybko,~K.; Buczko,~R.; Szczerbakow,~A.; Szot,~M.;
  \L{}usakowska,~E.; Balasubramanian,~T.; Wojek,~B.~M.; Berntsen,~M.~H.;
  Tjernberg,~O.; Story,~T. \emph{Nat. Mater.} \textbf{2012}, \emph{11}, 1023.


\setboolean{nobreakdashused}{false}\bibitem[Okada et~al.(2013)Okada, Serbyn,
  Lin, Walkup, Zhou, Dhital, Neupane, Xu, Wang, Sankar, Chou, Bansil, Hasan,
  Wilson, Fu, and Madhavan]{Okada13}
Okada,~Y.; Serbyn,~M.; Lin,~H.; Walkup,~D.; Zhou,~W.; Dhital,~C.; Neupane,~M.;
  Xu,~S.; Wang,~Y.~J.; Sankar,~R.; Chou,~F.; Bansil,~A.; Hasan,~M.~Z.;
  Wilson,~S.~D.; Fu,~L.; Madhavan,~V. \emph{Science} \textbf{2013}. \url{doi:
  10.1126/science.1239451}.


\setboolean{nobreakdashused}{false}\bibitem[Serbyn and Fu(2014)]{Serbyn14}
Serbyn,~M.; Fu,~L. \emph{Phys. Rev. B} \textbf{2014}, \emph{90}, 035402.


\setboolean{nobreakdashused}{false}\bibitem[Phuphachong
  et~al.(2017)Phuphachong, Assaf, Volobuev, Bauer, Springholz, de~Vaulchier,
  and Guldner]{Phuphachong17}
Phuphachong,~T.; Assaf,~B.~A.; Volobuev,~V.~V.; Bauer,~G.; Springholz,~G.;
  de~Vaulchier,~L.-A.; Guldner,~Y. \emph{Crystals} \textbf{2017}, \emph{7}, 29.


\setboolean{nobreakdashused}{false}\bibitem[Volkov and
  Pankratov(1985)]{Volkov85}
Volkov,~B.~A.; Pankratov,~O.~A. \emph{Sov. Phys. JETP} \textbf{1985},
  \emph{42}, 178.


\setboolean{nobreakdashused}{false}\bibitem[Korenman and
  Drew(1987)]{Korenman87}
Korenman,~V.; Drew,~H.~D. \emph{Phys. Rev. B} \textbf{1987}, \emph{35}, 6446.


\setboolean{nobreakdashused}{false}\bibitem[Agassi and
  Korenman(1988)]{Agassi88}
Agassi,~D.; Korenman,~V. \emph{Phys. Rev. B} \textbf{1988}, \emph{37}, 10095.


\setboolean{nobreakdashused}{false}\bibitem[Pankratov(1990)]{Pankratov90}
Pankratov,~O.~A. \emph{Semicond. Sci. Technol.} \textbf{1990}, \emph{5}, S204.


\setboolean{nobreakdashused}{false}\bibitem[Kolesnikov and
  Silin(1997)]{Kolesnikov97}
Kolesnikov,~A.~V.; Silin,~A.~P. \emph{J. Phys.: Condens. Mat.} \textbf{1997},
  \emph{9}, 10929.


\setboolean{nobreakdashused}{false}\bibitem[Kriechbaum(1986)]{Kriechbaum86}
Kriechbaum,~M. Envelope Function Calculations for Superlattices. In
  \emph{Two-Dimensional Systems: Physics and New Devices}; Bauer,~G.,
  Kuchar,~F., Heinrich,~H., Eds.;
\newblock Springer: Berlin, 1986;
\newblock p 120.


\setboolean{nobreakdashused}{false}\bibitem[Ando and Fu(2015)]{Ando15}
Ando,~Y.; Fu,~L. \emph{Annu. Rev. Condens. Matter Phys.} \textbf{2015},
  \emph{6}, 361.


\setboolean{nobreakdashused}{false}\bibitem[Dom{\'i}nguez-Adame(1994)]{Adame94}
Dom{\'i}nguez-Adame,~F. \emph{phys. stat. sol. (b)} \textbf{1994}, \emph{186},
  K49.


\setboolean{nobreakdashused}{false}\bibitem[Zhang et~al.(20125)Zhang, Kane, and
  Mele]{Zhang12}
Zhang,~F.; Kane,~C.~L.; Mele,~E.~J. \emph{Phys. Rev. B} \textbf{20125},
  \emph{86}, 081303(R).


\setboolean{nobreakdashused}{false}\bibitem[Tchoumakov et~al.(2017)Tchoumakov,
  Jouffrey, Inhofer, Bocquillon, Plaçais, Carpentier, and Goerbig]{Goerbig17}
Tchoumakov,~S.; Jouffrey,~V.; Inhofer,~A.; Bocquillon,~E.; Plaçais,~B.;
  Carpentier,~D.; Goerbig,~M.~O. \emph{Phys. Rev. B} \textbf{2017}, \emph{96},
  201302(R).


\setboolean{nobreakdashused}{false}\bibitem[Agassi(1994)]{Agassi94}
Agassi,~D. \emph{Phys. Rev. B} \textbf{1994}, \emph{49}, 10393.


\setboolean{nobreakdashused}{false}\bibitem[Pankratov et~al.(1987)Pankratov,
  Pakhomov, and Volkov]{Volkov87}
Pankratov,~O.~A.; Pakhomov,~S.~V.; Volkov,~B.~A. \emph{Solid State Commun.}
  \textbf{1987}, \emph{61}, 93.


\setboolean{nobreakdashused}{false}\bibitem[D{\'i}az-Fern{\'a}ndez
  et~al.(2017)D{\'i}az-Fern{\'a}ndez, Chico, Gonz{\'a}lez, and
  Dom{\'i}nguez-Adame]{Diaz-Fernandez17a}
D{\'i}az-Fern{\'a}ndez,~A.; Chico,~L.; Gonz{\'a}lez,~J.~W.;
  Dom{\'i}nguez-Adame,~F. \emph{Sci. Rep.} \textbf{2017}, \emph{8}, 8058.


\setboolean{nobreakdashused}{false}\bibitem[Assaf et~al.(2016)Assaf,
  Phuphachong, Volobuev, Inhofer, Bauer, Springholz, de~Vaulchier, and
  Guldner]{Assaf16}
Assaf,~B.~A.; Phuphachong,~T.; Volobuev,~V.~V.; Inhofer,~A.; Bauer,~G.;
  Springholz,~G.; de~Vaulchier,~L.~A.; Guldner,~Y. \emph{Sci. Rep.}
  \textbf{2016}, \emph{6}, 20323.


\setboolean{nobreakdashused}{false}\bibitem[Xu et~al.(2012)Xu, Liu, Alidoust,
  Neupane, Qian, Belopolski, Denlinger, Wang, Lin, Wray, Landolt, Slomski, Dil,
  Marcinkova, Morosan, Gibson, Sankar, Chou, Cava, Bansil, and Hasan]{Hasan12}
Xu,~S.-Y.; Liu,~C.; Alidoust,~N.; Neupane,~M.; Qian,~D.; Belopolski,~I.;
  Denlinger,~J.~D.; Wang,~Y.~J.; Lin,~H.; Wray,~L.~A.; Landolt,~G.;
  Slomski,~B.; Dil,~J.~H.; Marcinkova,~A.; Morosan,~E.; Gibson,~Q.; Sankar,~R.;
  Chou,~F.~C.; Cava,~R.~J.; Bansil,~A.; Hasan,~M.~Z. \emph{Nat. Comm.}
  \textbf{2012}, \emph{7}, 12505.


\setboolean{nobreakdashused}{false}\bibitem[D{\'i}az-Fern{\'a}ndez and
  Dom{\'i}nguez-Adame(2017)]{Diaz-Fernandez17b}
D{\'i}az-Fern{\'a}ndez,~A.; Dom{\'i}nguez-Adame,~F. \emph{Physica E}
  \textbf{2017}, \emph{93}, 230.


\setboolean{nobreakdashused}{false}\bibitem[Kriechbaum et~al.(1984)Kriechbaum,
  Ambrosch, Fantner, Clemens, and Bauer]{Kriechbaum84}
Kriechbaum,~M.; Ambrosch,~K.~E.; Fantner,~E.~J.; Clemens,~H.; Bauer,~G.
  \emph{Phys. Rev. B} \textbf{1984}, \emph{30}, 3394.


\setboolean{nobreakdashused}{false}\bibitem[Buczko and
  Cywi\ifmmode~\acute{n}\else \'{n}\fi{}ski(2012)]{Buczko12}
Buczko,~R.; Cywi\ifmmode~\acute{n}\else \'{n}\fi{}ski,~L. \emph{Phys. Rev. B}
  \textbf{2012}, \emph{85}, 205319.


\setboolean{nobreakdashused}{false}\bibitem[D\'{\i}az-Fern\'{a}ndez
  et~al.(2017)D\'{\i}az-Fern\'{a}ndez, Chico, and
  Dom\'{\i}nguez-Adame]{Diaz-Fernandez17c}
D\'{\i}az-Fern\'{a}ndez,~A.; Chico,~L.; Dom\'{\i}nguez-Adame,~F. \emph{J.
  Phys.: Condens. Matter} \textbf{2017}, \emph{29}, 475301.


\setboolean{nobreakdashused}{false}\bibitem[Feynman and
  Gell-Mann(1958)]{Feynman58}
Feynman,~R.~P.; Gell-Mann,~M. \emph{Phys. Rev.} \textbf{1958}, \emph{109}, 193.


\setboolean{nobreakdashused}{false}\bibitem[Dom\'{\i}nguez-Adame(1991)]{Adame91}
Dom\'{\i}nguez-Adame,~F. \emph{EPL (Europhysics Letters)} \textbf{1991},
  \emph{15}, 569.


\setboolean{nobreakdashused}{false}\bibitem[Glasser and Nieto(2015)]{Glasser15}
Glasser,~M.~L.; Nieto,~L.~M. \emph{Can. J. Phys.} \textbf{2015}, \emph{93},
  1588.


\end{thebibliography}

\vspace{3cm}
This article is published in full length in \textit{Beilstein J. Nanotech.}
\textbf{2017}, \textit{X}, No. Y.

\end{document}